# Improvement of all-optical Compton γ-rays source by reshaping colliding pulse

Q. Yu[1*], Y. Zhang[1†], Q. Kong[2‡] and S. Kawata[3]

[1] School of Mechanical Engineering and Rail Transit, Changzhou University, Changzhou 213164, China
[2] Institute of Modern Physics, Fudan University, Shanghai 200433, People's Republic of China
[3] Graduate School of Engineering, Utsunomiya University, Utsunomiya 321-8585, Japan

Abstract

All-optical Compton scattering is a remarkable method of generating high-quality γ radiation source. It is easier achieved in experiment by employing a pulse based on laser wakefield accelerator. The driving laser is backward reflected when wakefield acceleration stage is over and thus it naturally collides with energetic electrons. To increase reflected pulse intensity, parabolic focusing plasma mirror instead of flat reflecting target is usually adopted. But concave focusing mirror also deteriorates the emitted photon beam monochromaticity and collimation. We propose using stepped focusing plasma mirror to reflect the driving pulse to conquer these issues. The longitudinal length of reflected pulse by stepped target is larger and intensity is relatively small. It leads emitted photon beam to have better monochromaticity and collimation except for having larger emitted energy and higher laser utilization efficiency. We affirm the robustness of stepped focusing mirror reflecting regime through various kinds of numerical simulations.

Corresponding Author: *qin_yu_1@163.com; †zy@cczu.edu.cn
‡qkong@fudan.edu.cn

Based on the laser plasma interaction, researchers have proposed several methods to achieve high-quality, multi-MeV $\gamma$ radiation source. Such as, accelerated electrons transversely oscillate along with laser propagation in laser wakefiedl acceleration, in which energetic photon source can be generated through Beta oscillations of electrons [1]. One also can gain photon source via the generation of high-order harmonic wave in radiation pressure-dominated scheme [2]. The $\gamma$ ray can be obtained via homochromous radiation emission base on laser-solid interaction as well [3]. Another manner to achieve high-quality photon beam is bremsstrahlung during the interaction between atomic nucleus and hot electrons heated by laser [4]. In addition, energetic photons can be emitted in the processes of Compton scattering [5-7] and positron annihilation [8].

Recently, the generation of all-optical controllable $\gamma$ source has attracted widespread attention, attributed to the experimentally successful achievement of GeV-level electron beam [9-11]. Researchers have preliminary obtained high-quality, brilliant x/$\gamma$ rays [12-18] via all-optical Compton scattering. Generally, two methods attribute to realize all optical Compton scattering. In one way, two laser pulses are employed. One of them is utilized to evoke laser wakefield and inject background electrons into the wakefield to be accelerated and thus getting energetic electron beam. The other pulse propagating in the opposite direction collides with the accelerated electrons from wakefield when the acceleration stage is over. The energetic $\gamma$ photon bunch of high-brilliance with good collimation is produced in the colliding of laser and accelerated electrons via Compton scattering. The difficulty in experiment is the controls of alignment and temporal synchronization of the two counter-propagating lasers. To overcome these shortcomings, some groups present another ways to achieve all-optical Compton scattering employing only one pulse[13, 17]. The pulse evokes plasma wave and accelerates the background electrons. A solid target is placed in the front of the driving pulse after the electron acceleration finishes and the remaining laser is backward reflected by the solid target. The reflected laser naturally overlaps and collides with the energetic electrons to realize Compton scattering and produce a high-quality photon bunch[19].

The obtained photon energy $E_\gamma$ from all-optical Compton scattering is mainly ruled by three factors: the Lorentz factor of scattered electron beam $\gamma_0$, laser photon energy $\hbar\omega_L$ and normalized laser intensity $a_0$. The relation among emitted photon energy and above three factors is [20]: $E_\gamma \approx 4\gamma_0^2 \hbar\omega_L f(a_0)$ where $4\gamma_0^2$ is the relativistic Doppler frequency shifting. When colliding laser intensity is smaller, that is, $a_0 \ll 1$, $f(a_0) \approx 1$. If laser intensity is larger and satisfies $a_0 \geq 1$, $f(a_0) \approx a_0$. So emitted photon energy in all-optical Compton scattering is mainly limited by scattered electron energy and colliding laser intensity. We investigate the dependence of emitted photon energy on them by numerous numerical simulation cases and the results are displayed in Fig. 1(a). In the numerical simulations, a forward propagating energetic electron beam is presupposed and a counterpropagating laser overlaps and collides with the energetic electron beam. In all cases, simulative parameters are the same except for electron energy or normalized laser intensity. Both QED effect and Radiative reaction effect need to be considered if scattered electron energy and laser intensity are enough large in above Compton scattering process. The nonlinear QED effect can be quantified by the factor [21]: $\chi = 2\frac{\hbar\omega_L}{mc^2}\gamma_0 a_0$ where $m$ is the electron mass, c is the light speed. QED effect should be counted when $\chi \geq 1$ and QED effect can be ignored if $\chi \ll 1$. The black line in Fig. 1(a) exhibits the case of $\chi = 2\frac{\hbar\omega_L}{mc^2}\gamma_0 a_0 = 0.5$. In the reference system where particles are immobile, if $\lambda \gg \alpha\lambda_C$ and $E \ll E_{crit}/\alpha$, radiative reaction force is far less than Lorentz force and can be ignored. When it is in intense relativistic case, the radiative reaction force is comparable to Lorentz force and the radiative action force should be considered in this case. Di Piazza *et al.*, show the classical expression of radiative reaction force, namely, Landau-Lifschitz (LL) equation and give in the parameter range where radiative reaction force is dominant during laser electron colliding [22]: $R_c = \frac{2}{3}\alpha \frac{\hbar\omega_0}{mc^2}\gamma_0 a_0^2(1+\beta_x)$ where $\beta_x = \sqrt{\gamma_0^2-1}/\gamma_0$. Radiative reaction force is remarkable if $R_c \geq 1$ and radiative reaction force can be ignored when $R_c \ll 1$. The blue line in Fig. 1(a) displays the case of $R_c = \frac{2}{3}\alpha \frac{\hbar\omega_0}{mc^2}\gamma_0 a_0^2(1+\beta_x) = 0.5$. Fig. 1 seems to demonstrate two issues. One of them is that emitted photon energy always enlarges with the increase of electron energy with invariable laser intensity. The other is that emitted photon energy does not always enlarge along with the increase of laser intensity when electron energy is constant and there is a critical value of laser intensity. If laser intensity is smaller than the critical value, emitted photon energy increases along with the laser intensity. The emitted photon energy reaches saturation when laser intensity is above the critical intensity. Above critical intensity, the laser is enough intense to almost transfer entire electron beam energy into emitted photon source and thus it is invalid to furtherly improve emitted photon energy via persistent enhancing laser intensity. To furtherly prove the two points, we present the dependences of emitted photon energy on electron energy with some constant laser intensities of $a_0=200$, $a_0=250$ and $a_0=300$ in Fig. 1(c). All lines demonstrate emitted photon energy indeed enhances along with electron energy enlargement independent of laser intensity. When electron energies are constant $E_e=3\text{GeV}$, $E_e=5\text{GeV}$, $E_e=7\text{GeV}$, the relations between emitted photon energy and laser intensity are revealed in Fig. 3(b). It states emitted photon energy firstly enhances along with laser intensity when laser intensity is smaller than the critical intensity and then it reaches saturation when laser intensity is above the critical point. All results shown in Figs. 2(c) and (b) prove above analyses.

Two schemes should be implemented in order to increase emitted photon energy during Compton scattering process. One is to improve scattered electron energy. Laser wakefield accelerator provides an efficient way of generating energetic electron beams. Researchers have sufficiently investigated laser wakefield regime since the propose of laser wakefield accelerator [23] and the experimentally achievements of quasi-monoenergetic electron beams [24, 25] based on wakefield acceleration regime. They provide numerous methods to increase the accelerated electron energy both in theory [26-30] and experiment [9-11, 31, 32]. In theory, S. Yu. Kalmykova *et al.,*[26] investigated the injection, trapping, and acceleration of electrons in a three-dimensional nonlinear laser wakefield and achieved 1GeV electron beam of a few percent energy spread, moderate angular spread, and high charge (>100 pC per shot) employing a PW laser based on optimized LWFA. Laser wakefield acceleration at highly relativistic laser intensities was numerically and theoretically studied by utilizing three-dimensional particle-in-cell simulations in A. Pukhov's work [27]. They pointed out that 12-J, 33-fs laser pulses may produce bunches of $3 \times 10^{10}$ electrons with around 300MeV peak energy. Q. Yu *et al.*, extended electron dephasing length [28] and achieved multi-stage electron acceleration scheme [29] by employing layered gas targets with increasing density profiles. They also proposed a scheme [30] to improve accelerated electron beam collimation and emittance by adopting near-axis injection in a gas target with linearly increasing density profile. Many evident progresses have been made in experiment as well. R. Weingartner *et al.,* [31] gained ultralow emittance electron beams of a $0.21^{+0.01}_{-0.02}\pi$ mm mrad normalized transverse emittance in experiment based on a laser wakefield accelerator. In reference [9] the combination of self-guided laser propagation and ionization-induced injection has been achieved by which, the researchers accelerated electrons up to 1.45GeV energy in a laser wakefield accelerator. The 1 GeV barrier has also been surmounted in experiment in reference [10]. The investigators produced ultra-collimated monochromatic electron beams peaked at 2GeV. W. P. Leemans *et al.*, [11] experimentally generated 4.2GeV electron beams of 6% rms energy spread, 6 pC charge, and 0.3 mrad rms divergence in quick succession from a 9-cm-long capillary discharge waveguide with a plasma density of $\approx 7 \times 10^{17}$ cm$^{-3}$, powered by 0.3PW laser pulse. The emittance and brilliance of electron beams were furtherly experimentally improved in reference [32]. The other regime of increasing emitted photon energy from Compton scattering is improving colliding laser intensity. It can be accomplished with the aid of advanced high-power laser facilities. But, if we want to achieve above-mentioned all-optical Compton scattering utilizing only one pulse based on laser wakefield accelerator, the colliding laser is the reflected pulse of driving laser. The colliding pulse intensity, or say, the driving pulse intensity is limited to be small by the wakefield acceleration regime in order to maintain the stability of plasma wave profile for a long time.

To improve reflected laser pulse intensity during single-pulse all-optical Compton scattering process based on laser wakefield acceleration, some groups propose utilizing concave focusing plasma mirror replacing flat plasma mirror to reflect driving laser of wakefield as claimed in Fig. 2. It is the schematic diagram of single-pulse all-optical Compton scattering process employing concave focusing plasma mirror based on laser wakefield acceleration. The employed single laser drives wakefield to inject and accelerate electrons. When electron acceleration stage is over, the driving pulse strikes the concave focusing target. The focusing mirror focuses and reflects the driving pulse and the reflected pulse collides with accelerated electrons. The reflected pulse intensity greatly enlarges and radius evidently decreases as depicted in Fig. 2(b) where a free laser with intensity of $a_0 = 5$ is focused and reflected by a parabolic focusing plasma mirror as illustrated in Fig. 2(a). The reflected laser intensity reaches about $a_0 = 150$ through focusing effect of concave plasma mirror. The radius of reflected pulse prominently decreases due to the focusing effect as well. The reduction of laser radius leads more laser energy to take part in Compton scattering and thus improves laser utilization efficiency. If the reflecting target is flat mirror, the reflected laser radius is generally much larger than scattered electron beam radius, which is restricted by laser wakefield acceleration theory. So, only the middle part of laser participates in the Compton scattering and thus laser utilization efficiency is much smaller in flat reflecting target case.

But, some issues needing to be overcome still appear in all-optical Compton scattering based on parabolic concave focusing mirror. Firstly, ultra-tightly focusing effect leads reflected laser to diverge rapidly. The rapid divergence reduces the colliding time between electrons and reflected laser at the focusing position and thus lowers laser utilization efficiency. Secondly, the strong transverse electric field of ultra-tightly focusing reflected laser increases transverse momentum of electrons and thereby weakens the collimation of emitted photon bunch. Therefore, we here propose using focusing target of stepped profile as shown in Fig. 3(c) instead of parabolic profile to focus and reflect the driving pulse to avoid these disadvantages. Three kind of reflecting target structures including flat target, parabolic focusing target and stepped focusing target and the laser propagation path before and after being reflected by the three targets are plotted in Fig. 3. According to theoretical analyses, the lasers both in parabolic focusing mirror and stepped focusing mirror cases focus, and the reflected pulses are modulated in space and time by both targets. In former case, reflected pulse converges at the focusing point which leads the longitudinal length of reflected pulse to decrease. In latter case, the reflected laser converges at different points along with laser axis which causes longitudinal length of reflected pulse to increase compared with the former case.

To better prove above theoretical analyses and prediction, in the following, we compare the three all-optical Compton scattering based on wakefield acceleration scheme employing flat target (FT), parabolic focusing target (PT) and stepped focusing target (ST) via numerical simulations. The all-optical Compton scattering processes based on wakefield acceleration are hinted by Fig. 2 (only parabolic focusing target case is displayed) or Figs. 3(a)-(c) where wakefield acceleration stage is omitted and energetic electron beams from laser wakefield accelerator are presupposed. In the three scattering cases, flat reflecting target, parabolic focusing target and stepped focusing target are adopted. A forward-propagating energetic electron beam is presupposed in left and reflecting target locates in right. In the middle position, it is a free laser forward propagating as a driving pulse of wakefield. The driving laser is reflected by plasma mirrors of different structures and naturally overlaps with the counter propagating electron beam as rendered in Figs. 3(a)-(c). The simulations were performed based on PIC (Particle-in-Cell) code VLPL (Virtual Laser

Plasma Lab) [33]. The simulation parameters are the same in all cases unless specified. Simulation boxes are $70\lambda \times 70\lambda$, $120\lambda \times 70\lambda$ and $200\lambda \times 70\lambda$ for FT, PT, ST cases in $x$ and $y$ directions with resolution of $0.03\lambda \times 0.06\lambda$. The length and diameter of presupposed energetic electron bunch are $8\lambda$ and $1.2\lambda$ with density of $0.007n_c$. There are 8 macro-particles in each cell to standing for energetic electron beam. The electron beam propagates from left to right with 0.3GeV peak energy and 5% energy spread. There are 8 macro-particles in each cell to hint reflecting targets and their densities are $5n_c$. For flat target case, the reflecting mirror is flat with thickness and height of $5\lambda \times 70\lambda$. For PT case, the reflecting plasma mirror is concave with parabolic profile of $x = ky^2$ where $k \approx 0.04$. The parabolic vertex lies in $x = 90\lambda$ and the thickness of parabolic focusing target along with laser propagating axis is $15\lambda$. In stepped focusing target case, the reflecting mirror profile can be expressed as:

$$x = \begin{cases} x_1 - \xi y^2, & y_4 < |y| < y_5 \\ x_2 - \xi y^2, & y_3 < |y| < y_4 \\ x_3 - \xi y^2, & y_2 < |y| < y_3 \\ x_4 - \xi y^2, & y_1 < |y| < y_2 \\ x_5 - \xi y^2, & 0 < |y| < y_1 \end{cases} \text{ where } x_1=155, x_2=165, x_3=175, x_4=185,$$

$x_5=195$, $y_1=4$, $y_2=10$, $y_3=17$, $y_4=25$, $y_5=35$ and $\xi=0.006$. The laser is linearly polarized with wavelength and radius of $\lambda = 1\mu m$, $R = 20\mu m$. Normalized laser intensity is $a_0 = 0.8$ and the half pulse duration is $20\lambda$.

We exhibit the evolutions of reflected laser profiles with time for PT and ST cases in Fig. 4 and the laser outline is almost invariable before and after being reflected in FT case. In PT case, laser focuses rapidly after reflected. The pulse longitudinal spatial length is very small and is only about $8\lambda$ as displayed in Fig. 4(c) and (d), because the reflected pulse converges at a point as predicted by above-mentioned theory in Figs. 3(b) and (e). The most laser energy is reflected to very small space. So the reflected laser is more intense and its maximum is up to $a_0 = 3$. In ST case, the longitudinal spatial length of reflected pulse is much larger and it reaches approximate $70\lambda$ as revealed in Figs. 4(h) and (i), because the reflected pulse converges at different longitudinal points as demonstrated in Figs. 3(c) and (f). Even though it also tightly focuses in ST case, the laser energy is reflected to larger space, so the reflected laser intensity is relatively small and the maximum intensity is $a_0 = 2$.

The reflected pulses naturally collide with presupposed energetic electron beams for three cases. In PT and ST cases, the colliding happens when reflected laser focusing process just finishes. To display the colliding process between reflected laser and electrons, we draw the temporal evolutions of total electron energy $E_{e_{total}}$, total emitted photon energy $E_{p_{total}}$ and emitted photon number $N_p$ for the three cases in Fig. 5 where picture (a) and (d) express the situation of FT case, drawing (b) and (e) explain the circumstance of PT case and the particles energies hinted in figures (c) and (f) are from ST case. Electron energies reduce slowly and meanwhile emitted photon energies increase rapidly particularly at the most violent moments of colliding in all cases. After the colliding finishes, the total emitted photon energies reach $6.4 \times 10^{-7} J$, $1.89 \times 10^{-6} J$ and $3 \times 10^{-6} J$ in FT, PT and ST cases, and the photon numbers reach $3.25 \times 10^7$, $3.85 \times 10^7$ and $8.67 \times 10^7$ respectively. That is, laser utilization efficiency improves to 4.7 times and 1.6 times of flat target and parabolic target cases by our proposed stepped target. The $\gamma$ ray brightness likewise increases to 2.7 times and 2.3 times of FT and PT cases by proposed stepped target. Contrasted with FT case, parabolic target seems not to greatly increase emitted photon number. It is because the colliding time is relative short even though colliding pulse intensity is larger in PT case. The blue line in Fig. 5(b) reveals the colliding time is about $10\tau$ in PT case where $\tau$ is a laser period. The presented ST not only increases emitted energy but also enlarges emitted photon number contrasted with FT case. We can see the colliding in ST case can keep about $30\tau$ from the blue line of Fig. 5(c). The spatio-temporal modulation of reflected pulse caused by a stepped reflecting focusing target increases colliding pulse intensity and the longitudinal length. The larger longitudinal length prolongs colliding time of Compton scattering. The increases of both colliding laser intensity and colliding time guarantee high emitted energy and larger laser utilization efficiency in ST case.

The emitted photon energy can be estimated as following: for a linear or single photon scattering, considering energy and momentum conservations, the emitted photon frequency can be described as: $\omega_p \approx \frac{4\gamma_0^2 \omega_L}{1 + \theta^2\gamma_0^2 + \frac{2\gamma_0^2 \hbar \omega_L}{mc^2}}$. In electron scattering relativistic laser case, radiation reaction force and QED effect should be taken into account. In this case, the emitted photon frequency is modified as $\omega_p = \frac{4\gamma_0^2 n\omega_L}{1 + a_0^2 + 4\gamma_0 n\omega_L}$ according to Chris Harvey's [34] work. The focusing target enhances emitted photons energy and number via improving reflected laser intensity. It can be explained in the following two aspects. Firstly, The radiation reaction force [22] is described as: $f_{RR} \sim 4\pi\alpha\hbar\omega_L\gamma_0^2 a_0^2/(3\lambda)$. The increase of laser intensity causes radiation reaction force to enlarge and thus leads to the increase of radiation energy. Secondly, Nikishov and Ritus pointed out that $a_0^2$ is proportional to $E^2$, that is the density of laser photons $n_\gamma$. The detail relation between them is $a_0^2 = \frac{\hbar e^2}{m^2 c^2 \omega_L} n_\gamma = 4\pi\alpha\nu^2\lambdabar^3 n_\gamma$ where $\nu \equiv \hbar\omega_L/mc^2$ is normalized laser frequency and $\lambdabar^3 n_\gamma$ is the number of laser photon in a cube with the length of a laser wavelength. The probability of multi-photon scattering $e^- + n\gamma_L \rightarrow e^- + \gamma$ is proportional to $a_0^{2n} \sim n_\gamma^n$. In other words, the increase of laser intensity leads the probability of multi-photon Compton scattering to enlarge. So obtained emitted photon number increases when reflected laser intensity enlarges. Above two points demonstrate more intense the reflected laser, more energy and photons are emitted. Furthly, ST increases total emitted photon energy and number compared with PT case due to bigger longitudinal spatial length which leads to longer colliding time.

Finally, we get γ rays source after reflected laser scatters electrons for all cases. But, the photon beam qualities are different with different reflecting targets. We describe energy and angular spectra of emitted photon beams with different reflecting targets in Fig. 6 where (a) and (d) present the energy and angular spectra of FP case. Energy and angular spectra from PT and ST cases are shown in Figs. 6(b), (e) and Figs. 6(c), (f). The peak energy and energy spread are 0.125MeV and 31% in flat target case. The peak energy enlarges to 0.26MeV when flat target is replaced by parabolic focusing target, but the monochromaticity of photon bunch deteriorates and energy spread reaches up to 75%. Our proposed ST lowers the photon beam energy spread to 19% not at the cost of peak energy reduction compared with PT case and the peak energy is 0.252MeV. It is because the larger longitudinal spatial length of reflected pulse in ST case makes the electron beam to feel more even force in longitudinal direction. It leads to smaller energy spread of emitted photon bunch. From the three angular spectra as described in Figs. 6(d)-(f), one can see PT makes the photon beam angular divergence to increases compared with FT case. It is because focusing effect enhances transverse force felt by scattered electrons and thus increases emitted photons transverse momentum. Even though ST also causes laser focusing effect, the reflected laser longitudinal length is bigger which weakens the focused laser intensity. Consequently, the transverse momentum, angular divergence and collimation of emitted photons are improved in ST case compared with PT case.

To verify the robustness of stepped reflecting target scheme, we perform other simulations via adjusting electron beam parameters and stepped target parameters and other parameters remain, unless specified. The simulative results of adjusting the parameters of electron beams are shown in Fig. 7. Picture (a) and (b) display the evolutions of total emitted photon energy and photon number with time for FT, PT and ST cases when electron beam diameter is enlarged to $d = 2\mu m$. It is clearly claimed the PT enhances emitted photon energy to $2.67 \times 10^{-6}$ J from $1.06 \times 10^{-6}$ J of FT case. And meanwhile the emitted photon number likewise increases to $6.1 \times 10^7$ from $5.3 \times 10^7$. The ST furtherly enlarges emitted photon energy and number to $4.31 \times 10^{-6}$ J and $1.38 \times 10^8$. It well explains ST can improve emitted energy, photon number and laser utilization efficiency compared with both FT and PT cases. The temporal evolutions of emitted photon energy and number, with electron beam length being reduced to $L = 6\lambda$ are plotted for the three cases in Figs 7(c) and (d). The PT enhances both emitted photon energy and number and they are up to $1.42 \times 10^{-6}$ J and $2.9 \times 10^7$ from $0.49 \times 10^{-6}$ J and $2.5 \times 10^7$ of FT case. Then, the ST furtherly improves the emitted photon energy and number to $2.22 \times 10^{-6}$ J and $6.2 \times 10^7$. It indicates ST still works well when energetic electron beam longitudinally shortens.

In Fig. 8, we change the stepped target parameters to demonstrate the robustness of Compton scattering based on ST scheme. In Figs. 8(a) and (b), we employ the stepped target parameters of $x_1 = 159$, $x_2 = 168$, $x_3 = 177$, $x_4 = 186$, and $x_5 = 195$, and the temporal evolutions of emitted photon energy and number of three cases are revealed. Evidently, both emitted photon energy and number are increased by PT target and they reach $1.9 \times 10^{-6}$ J and $3.9 \times 10^7$ from $0.6 \times 10^{-6}$ J and $3.3 \times 10^7$ of FT case. The ST scheme further enhances emitted photon energy and number to $3.2 \times 10^{-6}$ J and $8.7 \times 10^7$. The ST parameters are $y_1 = 4$, $y_2 = 12$, $y_3 = 20$, $y_4 = 28$, and $y_5 = 35$ in Figs. (c) and (d). Similarly, we present the evolutions of emitted photon energy and number with time for three cases. The ST mechanism improves emitted photons energy up to $3.4 \times 10^{-6}$ J. It is 1.8 times and 5.7 times of PT and FT cases. It also improves emitted photon number at the same time. Now, there are $8.7 \times 10^7$ emitted photons in ST case. It increases to about 2.3 times and 2.6 times of PT and FT cases. Fig. 8 sufficiently illustrates that the ST scheme works well when utilizing other stepped target parameters. Figs. 7 and 8 demonstrate that ST mechanism works well with different electron beam and stepped target parameters. Of course, more parameters need to be considered in order to verify the stability of ST scheme and to find out the parameter range where ST evidently increases the energy and number of emitted photon source. In the following works, we will research the effects of all kind of parameters and give out the parameter range in which ST scheme reacts well.

In conclusion, we investigate all-optical Compton scattering achieved by a laser pulse which both evokes laser wakefield and is backward reflected by reflecting target via numerical simulations. In this work, we employ three kinds of reflecting targets including flat profile, parabolic focusing profile and stepped focusing profile and study the effects brought by different target profiles on Compton scattering. Simulative results reveal parabolic focusing plasma mirror efficiently enlarges reflected laser intensity and thus improves emitted $\gamma$ ray source energy compared with flat plasma mirror. But, the emitted photon number does not evidently enhance. What makes it terrible is the obtained $\gamma$ photon beam energy spread and collimation deteriorate due to ultra-tightly focusing effect caused by parabolic focusing plasma mirror. We propose using stepped focusing mirror to replace parabolic focusing mirror to reflect driving pulse. The stepped focusing target prolongs reflected pulse longitudinal spatial length and thus improves monochromaticity and collimation of emitted photon beam except for increasing emitted energy. With a variety of scattered electron beam and reflecting target parameters, the stepped focusing mirror works well.

**Acknowledgements**

This work is supported by Natural Science Foundation of China under Contract No. 11804348.

# Figure Captions

**Figure 1**

The energetic electron beams collide with counter-propagating laser pulses with various electron energies and laser intensities. The dependence of emitted photon beam energy on scattered electron energy and laser intensity in the process of Compton scattering is plotted in picture (a). The black line shows the case of $\chi = 0.5$ and the case of $R_c = 0.5$ is displayed by the green line. In figure (b), we present the dependence of emitted photon energy on laser intensity with constant scattered electron energy. The black, red and blue lines demonstrate the case of colliding electron energy $E_e = 3 GeV$, $E_e = 5 GeV$ and $E_e = 7 GeV$. The relation between emitted photon beam energy and electron beam energy is revealed in drawing (c). The black line demonstrates the situation of $a_0 = 200$ and meanwhile the red and blue lines exhibit the circumstances of $a_0 = 250$ and $a_0 = 300$, respectively.

**Figure 2**

The schematic diagram of all-optical Compton scattering based on wakefield accelerator and concave focusing plasma mirror is hinted in figure (a). Only one pulse is employed both as driving pulse and colliding pulse. The laser evokes laser wakefield and accelerates the injected electrons behind it. When wakefield acceleration stage is over, the driving laser is reflected by the focusing plasma mirror and it naturally overlaps with energetic electrons from laser wakefield accelerator to realize Compton scattering. We simulate the focusing process of reflected pulse of a forward-propagating free laser by parabolic focusing reflecting target. We plot laser transverse electric field distributions after focusing in (b). The concave plasma mirror successfully focuses laser intensity to be about $a_0 = 150$ from original $a_0 = 5$.

**Figure 3**

The reflecting target profiles and driving laser propagating paths during reflecting are pictured. Drawing (a) and (d) demonstrate the reflecting target profile and laser reflecting path for FT case. The reflecting target structure and laser propagation path before and after reflecting for PT case are plotted in (b) and (e). The stepped focusing plasma mirror profile is shown in (c). The reflecting path of driving laser by stepped focusing mirror is described in figure (c) and (f).

**Figure 4**

The reflecting target structures for PT and ST cases are depicted in figure (a) and (f), respectively. The reflected pulse profiles by PT are described in (b), (c), (d) and (e) at $t = 60\tau$, $t = 65\tau$, $t = 70\tau$ and $t = 85\tau$. The transverse electric field distributions of reflected laser in ST case are depicted in figure (g), (h), (i) and (j) for the moments of $t = 100\tau$, $t = 105\tau$, $t = 110\tau$ and $t = 140\tau$.

**Figure 5**

The temporal evolutions of total electron energy and emitted photon energy for flat target, parabolic target and stepped target cases are exhibited in (a)-(c). Figure (d)-(f) reveal the evolutions of generated photon number via Compton scattering in FT, PT and ST cases.

**Figure 6**

The obtained photon bunches energy and angular spectra for three cases are painted. Figure (a) and (d) present the energy and angular spectra of photon beam from FT case. The energy and angular spectra of photon beam in PT case are displayed in (b) and (e). The profiles of energy and angular spectra for ST case are demonstrated in (c) and (f).

**Figure 7**

The temporal evolutions of produced photons energy for FT, PT and ST cases are hinted by the black, red and blue lines in (a), when energetic electron beam diameter increases to $2\lambda$. The evolutions of obtained photon number with time for above three cases are depicted by black, red and blue lines in (b). In drawing (c), we plot the temporal evolutions of produced photon total energy for FT, PT and ST cases revealed by black, red and blue lines with scattered electron beam length decreasing to $6\lambda$. The photon number evolutions with time for FT, PT and ST cases with $6\lambda$ longitudinal electron bunch length are depicted by black, red and blue lines in (d).

**Figure 8**

The temporal evolutions of emitted photon bunch total energy in FT, PT and ST cases with the reflecting target parameters of $x_1 = 159$, $x_2 = 168$, $x_3 = 177$, $x_4 = 186$, and $x_5 = 195$ are shown in (a). The evolutions of photon number with time for FT, PT and ST cases are demonstrated via black, red and blue lines in (b). The energy evolutions of produced photon bunches with time for FT, PT and ST cases are depicted by black, red and blue lines in (c) where the stepped reflecting target parameters are modified to $y_1 = 4$, $y_2 = 12$, $y_3 = 20$, $y_4 = 28$, and $y_5 = 35$. The photon number evolutions with time for FT, PT and ST cases are pictured by black, red and blue lines in (d).

**Figure 1**

**Improvement of all-optical Compton γ-rays source by reshaping colliding pulse**

**Q. Yu** *et al.*,

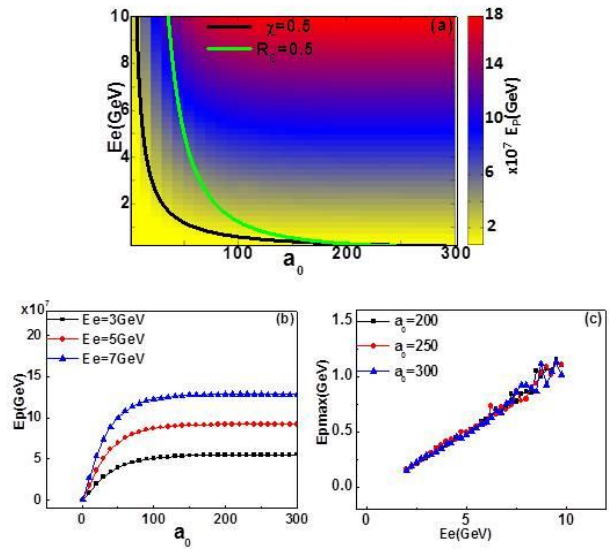

**Figure 2**

**Improvement of all-optical Compton γ-rays source by reshaping colliding pulse**

**Q. Yu *et al.*,**

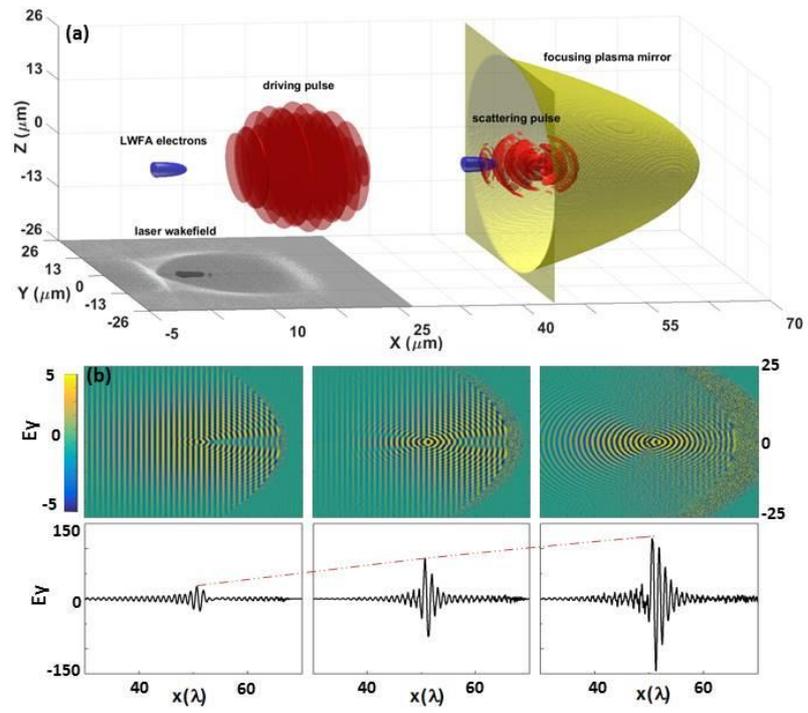

**Figure 3**

**Improvement of all-optical Compton γ-rays source by reshaping colliding pulse**

Q. Yu *et al.*,

Q. Yu

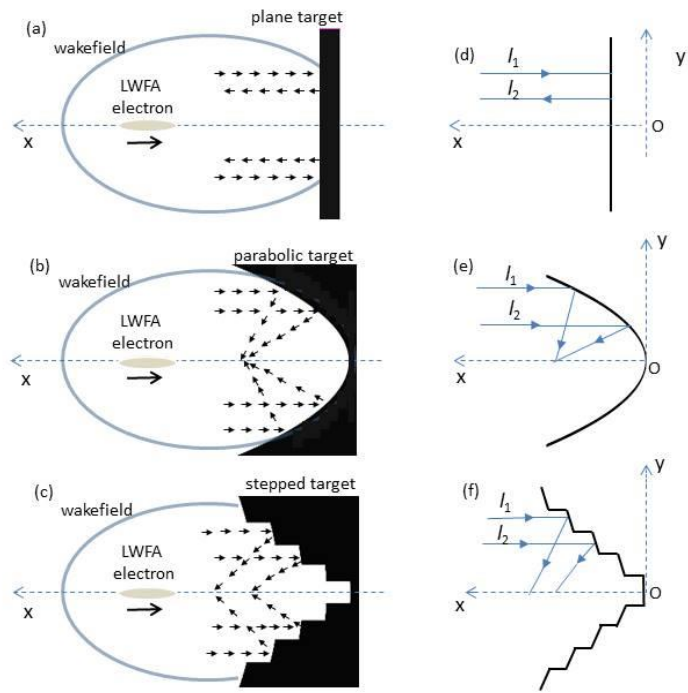

**Figure 4**

**Improvement of all-optical Compton γ-rays source by reshaping colliding pulse**

**Q. Yu *et al.*,**

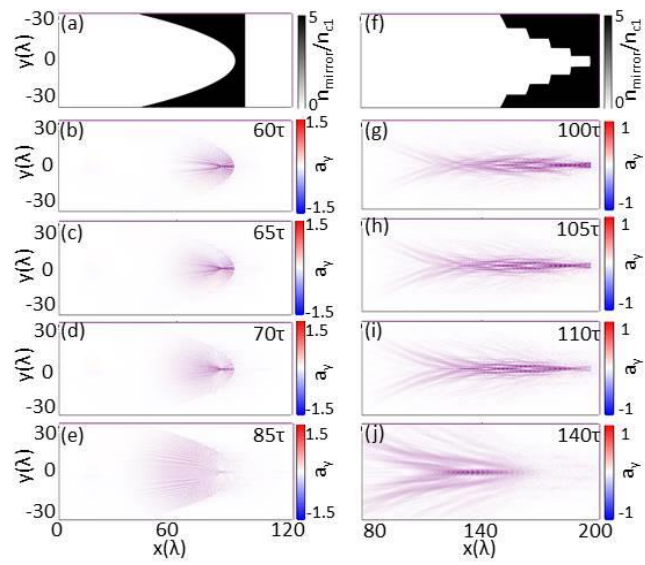

**Figure 5**

**Improvement of all-optical Compton γ-rays source by reshaping colliding pulse**

**Q. Yu *et al.*,**

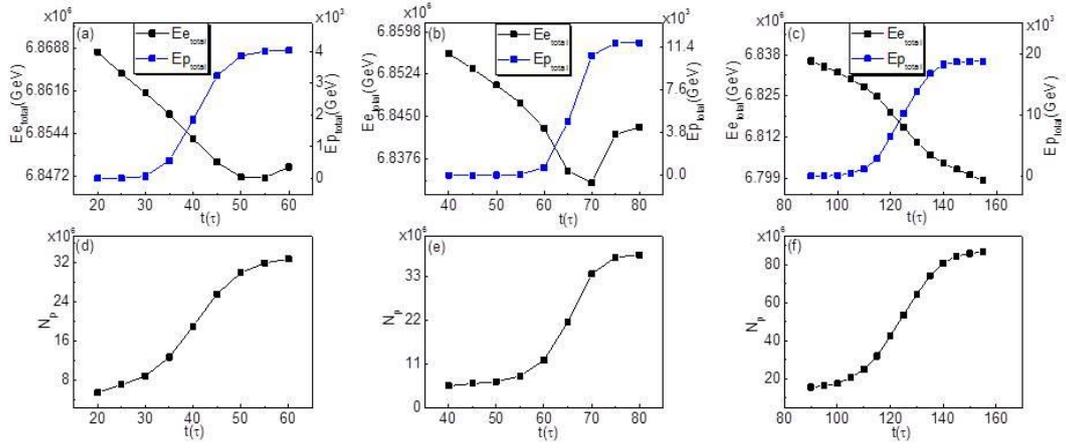

**Figure 6**

**Improvement of all-optical Compton γ-rays source by reshaping colliding pulse**

**Q. Yu *et al.*,**

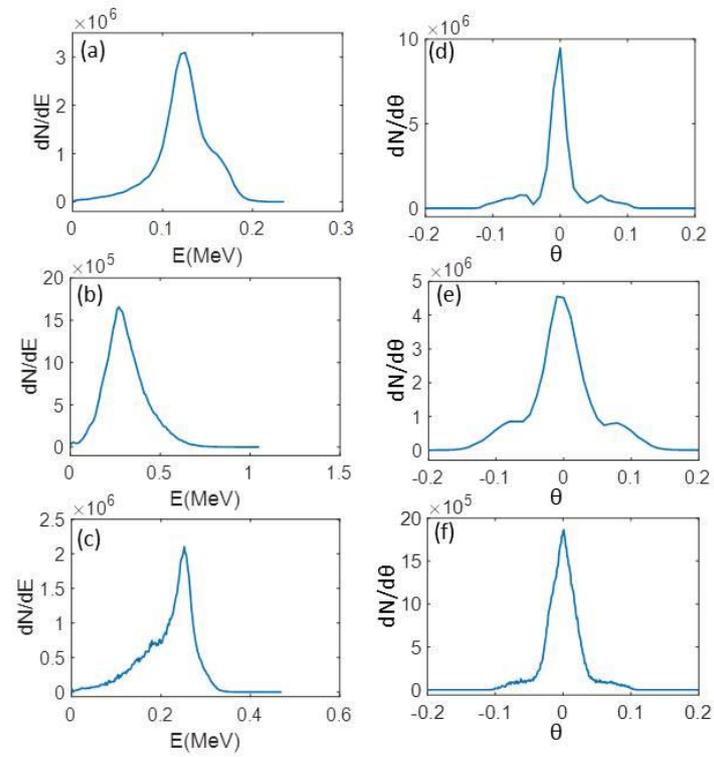

**Figure 7**

**Improvement of all-optical Compton γ-rays source by reshaping colliding pulse**

**Q. Yu *et al.*,**

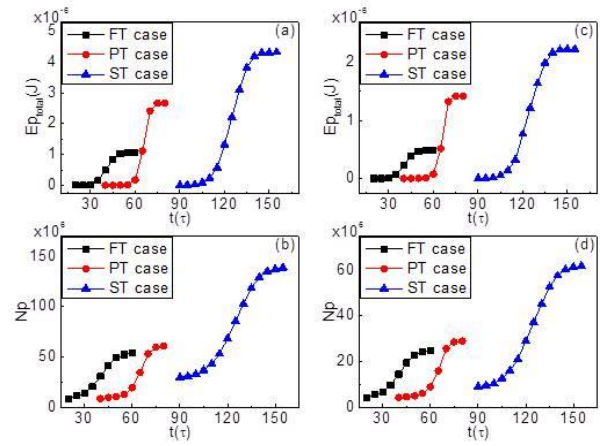

**Figure 8**

**Improvement of all-optical Compton γ-rays source by reshaping colliding pulse**

**Q. Yu *et al.*,**

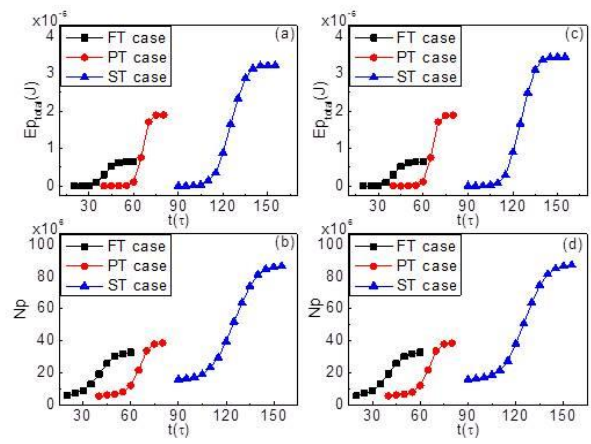